\documentclass[aps,prd,twocolumn,superscriptaddress,showpacs,floatfix,nobibnotes]{revtex4}

\usepackage{latexsym}
\usepackage{amsmath}
\usepackage{amssymb}
\usepackage{graphicx}
\usepackage{longtable}
\usepackage{bm}
\usepackage{epsfig}
\usepackage{citesort}
\usepackage{slashed}

\begin{document}

\title{Dyson-Schwinger equations with a parameterized metric}

\author{Wei Yuan}
\affiliation{Department of Physics and State Key Laboratory of
Nuclear Physics and Technology, Peking University, Beijing 100871,
China}
\author{Si-xue Qin}
\affiliation{Department of Physics and State Key Laboratory of
Nuclear Physics and Technology, Peking University, Beijing 100871,
China}
\author{Huan Chen}
\affiliation{Institute of High Energy Physics, Chinese Academy of
Science, Beijing 100049, China.}
\author{Yu-xin Liu}
\email[Corresponding author: ]{yxliu@pku.edu.cn}
\affiliation{Department of Physics and State Key Laboratory of
Nuclear Physics and Technology, Peking University, Beijing 100871,
China} \affiliation{Center of Theoretical Nuclear Physics, National
Laboratory of Heavy Ion Accelerator, Lanzhou 730000, China}

\date{\today}

\begin{abstract}
We construct and solve the Dyson-Schwinger equation (DSE) of quark
propagator with a parameterized metric, which connects the Euclidean
metric with the Minkowskian one. We show, in some models, the
Minkowskian vacuum is different from the Euclidean vacuum. The usual
analytic continuation of Green function does not make sense in these
cases. While with the algorithm we proposed and the quark-gluon
vertex ans\"{a}tz which preserves the Ward-Takahashi identity, the
vacuum keeps being unchanged in the evolution of the metric. In this
case, analytic continuation becomes meaningful and can be fully
carried out.
\end{abstract}


\maketitle

\parskip=3pt

\section{Introduction}

The true metric of space-time is certainly the Minkowskian other
than the Euclidean. However, with the Minkowskian metric, in many
circumstances, it is very difficult to perform calculations of
physical quantities of our interests. Such difficulties are mainly
caused by the kinematics which are only included in the Minkowskian
system but not in the Euclidean one. Nevertheless, in the framework
of path integrals in quantum field theory, the vacuums implied by
the Minkowskian metric (with an infinitesimal $\varepsilon$-term)
and the Euclidean metric should be the same (see, for example,
Ref.~\cite{Peskin-textbook} or section II of this paper). Therefore,
if we only focus on some static quantities such as density,
condensate and space-like Green function, Euclidean metric and
Minkowskian metric are equivalent, giving the same expectation
value. However, if we also have interests in the physical quantities
which involve in transport processes such as conductance and
viscosity, the Minkowskian metric would be inevitable. In recent
years, viscosity of strongly coupled quark-gluon plasma
(sQGP)~\cite{sQGP} and the problem of energy losing during quark jet
quenching process in sQGP have attracted a great deal of attention.
These problems all involve transport processes and require
Minkowskian metric to be implemented~\cite{Wang2007}.

How could we realize Minkowskian metric from an Euclidean system?
The most popular way is the analytic continuation (Wick rotation) if
we know the concrete analytic form of the Green function. We can
easily apply such an approach in perturbative calculations. However,
in many non-perturbative cases, for examples, the lattice QCD
calculations~\cite{lattice} and the Dyson-Schwinger equation (DSE)
approach~\cite{DSE} which are constructed in Euclidean space, we can
hardly perform an analytic continuation directly to the numerical
results calculated in Euclidean space. Of course we might use these
results to fit an analytic form in some
speculations~\cite{analytic1,Chen2008}. For instance, a useful
priori decomposition rules is the {\it K\"{a}ll\'{e}n-Lehmann}
spectral representation. However, the problem arises from the lack
of a systematic step-by-step program to efficiently assess and
improve the accuracy of these speculations.

In many models built in Euclidean space, another even more serious
problem bothering us is whether the analytic continuation of
Euclidean Green function is meaningful and self-consistent with the
principle of quantum field theory, namely whether the Minkowskian
path integral formalism and the Euclidean one lead to the same
vacuum.

In this paper, we will develop a new scheme, in the framework of
DSE, which is free of all the problems mentioned above. The paper is
organized as follows. At first, in section II, we introduce the
parameterized metric in the context of Quantum Field Theory (QFT)
and show how it works as an effective parameter for the task of
analytic continuation. In section III, the so-called vacuum problem,
which bothers us in some models, is described explicitly. Then, we
construct Dyson-Schwinger equation for quark with the parameterized
metric in section IV. The numerical algorithm and calculated results
are given in section V. Finally, in section VI, we make a summary
and give some remarks.

\section{parameterized metric and its validity in some simple cases}

To introduce a parameterized metric, we should deal with the metric
as a variable everywhere in QFT. For example, we consider the simple
$\phi^4$ theory. Its action can be expressed as
\begin{equation}
S=\int d^{4}x^{\mu}
\sqrt{-\text{det}g_{\mu\nu}}\{-\frac{1}{2}g^{\mu\nu}
\frac{\partial \phi}{\partial x^{\mu}}\frac{\partial \phi}{\partial x^{\nu}}
-\frac{1}{2}m^{2}\phi^{2}-\frac{\lambda}{4!}\phi^{4}\}.
\end{equation}
The covariant Minkowskian/Euclidean metric adopted in this paper is
$g^M_{\mu \nu}=diag(-1,+1,+1,+1)$ and $g^E_{\mu
\nu}=diag(+1,+1,+1,+1)$ and the contravariant metric satisfies
$g^{\mu \lambda} g_{\lambda \nu}=\delta^{\mu}_{\nu}$. Considering
the requirements of $\sqrt{-\text{det}g^{E}_{\mu\nu}}=-i$ and
$\sqrt{-\text{det}g^{M}_{\mu\nu}}=1$, we could parameterize the
covariant metric as
\begin{equation} g_{\mu\nu}(\varepsilon)= \left(
\begin{array}{cccc}
 e^{i\varepsilon} & 0 & 0 & 0 \\
 0 & 1 & 0 & 0 \\
 0 & 0 & 1 & 0\\
 0 & 0 & 0 & 1
\end{array}
\right)\, ,
\end{equation}
with $\varepsilon \in [0,\pi)$. Then we have $g(\varepsilon=0)
\equiv g^E$ and $g(\varepsilon \rightarrow \pi)\rightarrow g^M$.
Furthermore, one could easily prove that $iS_{M}[\phi] = i
S[g(\pi),\phi]$ and $-S_{E}[\phi] = i S[g(0),\phi]$, so that the
uniform path-integral formalism reads
\begin{equation}
\int [d\phi]e^{iS[g(\varepsilon),\phi]} \, .
\end{equation}

Up to now, we have only found a simple way (namely, varying the
parameter $\varepsilon$) to connect $g_{M}$ with $g_{E}$. In fact,
there are infinite ways to realize this connection. Why ours is
preferred, or what does it mean? To answer these questions, as an
simple example, we construct the quantum theory of the $\phi^4$
model from the Lagrangian
$$
L=\sqrt{-\text{det}g_{\mu\nu}(\varepsilon) } \{-\frac{1}{2}
g^{\mu\nu}(\varepsilon) \frac{\partial \phi}{\partial
x^{\mu}}\frac{\partial \phi}{\partial x^{\nu}}
-\frac{1}{2}m^{2}\phi^{2}-\frac{\lambda}{4!}\phi^{4}\}.
$$
We will use the Hamiltonian formulation to quantize it. The momentum
density conjugate to $\phi(x)$ is
\begin{equation}\label{doliang}
\pi(x)\equiv\frac{\partial L}{\partial\dot{\phi}} = -
g^{00}(\varepsilon) \sqrt{-\text{det}g_{\mu
\nu}(\varepsilon)}\,(\dot{\phi}).
\end{equation}
By using the relation $\text{det}g_{\mu \nu}=g_{00}=\frac{1}{g^{00}}$,
we obtain the Hamiltonian
\begin{equation}
H\equiv\int d^{3}x [\pi \dot{\phi}-L]\! = \! e^{i(\varepsilon\! -
\!\pi)/2} \! \int
d^{3}x[\frac{1}{2}\pi^{2}+\frac{1}{2}m^2\phi^2+\frac{\lambda}{4!}\phi^4].
\end{equation}
At $ t = 0 $, we suppose that $\phi(\vec{x})$ and $\pi(\vec{x})$ are
Hermite operators and satisfy the relation
\begin{eqnarray}\label{duiyi}
\begin{split}
[\phi(\vec{x}),\pi(\vec{y})] & =  i\delta^{(3)}(\vec{x}-\vec{y}) \,
, \\ [\phi(\vec{x}),\phi(\vec{y})] & =
[\pi(\vec{x}),\pi(\vec{y})]=0\, ,
\end{split}
\end{eqnarray}
and the equation of motion for Heisenberg operator is
\begin{equation}\label{eom}
O(t)=U(t)^{-1}O(0)U(t),
\end{equation}
where the evolution matrix is
\begin{equation} \label{ut}
U(t)=e^{-iHt}=e^{-iH_{M}\tilde{t}}.
\end{equation}
Here, $H_{M}$ is the Hamiltonian operator in Minkowskian space as
the same as that commonly used, and $\tilde{t} \equiv
e^{i(\varepsilon-\pi)/2} t$. We should note that $U(t)$ is not a
unitary matrix since the $H$ is not Hermite. So the operators
$\phi(t)$ and $\pi(t)$ need not to be Hermite when $t\neq0$, but it
does not affect Eq.~(\ref{duiyi}) to be held at $t\neq0$. In fact,
if $\phi(t=0)$ is Hermite, from Eqs.~(\ref{doliang}), (\ref{eom})
and (\ref{ut}) we can prove that $\pi(t=0)$ is also Hermite (note
that $\dot{\phi}(0)$ is not Hermite). Moreover, one could easily
prove that the correlation function G(x,y) in vacuum is
time($t$)-translation invariant. From these observations, we
conclude that our construction is self-consistent.

In addition, there are some other properties that we need for our
purpose in this paper. Firstly, the structure of eigenstates of the
$H$ is independent of $\varepsilon$. Secondly, the effects of the
parameterized metric we adopt is equivalent to
%
%
the Wick rotation to time axis in complex plane but nothing else. In
fact, one prefers to regard Eq.~(\ref{ut}) as the standard
definition of Wick rotation and take it to realize analytic
continuation of Green functions. However, in path-integral
formalism, our method is more convenient. As we know that the
path-integral formalism can automatically perform the time order
operator, but the in/out states are unfixed in it. Nevertheless, the
parameterized metric
%
%
undertakes an extra responsibility, besides the Wick rotation, that
guarantees the in/out states to be those in vacuum (the lowest
energy state of $H_M$). It is true in our method with varying
$\varepsilon \in [0,\pi)$. To see it more explicitly, we just need
to show that
\begin{equation} \label{fv}
U(\infty)|\alpha\rangle\propto e^{-H_{M}cos(\varepsilon/2)\infty}
\mid \alpha>\rightarrow |vac\rangle \, .
\end{equation}
It should be emphasized that Eq.~(\ref{fv}) is definitely true
except for the case with $\varepsilon=\pi$, so that we could only
take the limit $\varepsilon \rightarrow \pi$ to realize Minkowskian
space whose effects are equivalent to the $\varepsilon$-term trigger
adopted commonly (see, for example, Ref.~\cite{Peskin-textbook}).
Until now, we could say that this parameterized metric, used in
path-integral formalism for analytic continuation of Green
functions, is legitimate. (In addition, Eq.~(\ref{fv}) means that
the vacuum state implied by $g(\varepsilon)$ is identical and
independent of $\varepsilon$. However, as we will see in the next
section, such a statement is not always satisfied by models. We
refer it as vacuum problem.)

As a simple test of our method, we implement it to calculate the
one-loop electron self-energy. At first, we need conventions to
simplify our representation. In this paper, real space-time
variables are always thought to be the components in a contravariant
vector $x^{\mu}$. Naturally, the real momentum variables are always
thought to be those in a covariant vector $p_{\mu}$. The Fourier
transform is
\begin{equation}
f(x^{\mu})=\int \frac{d^{4}p_{\mu}}{(2\pi)^{4}}
\sqrt{-\text{det}g^{\mu\nu}(\varepsilon)}e^{-ip_{\mu}x^{\mu}}f(p_{\mu})\,
,
\end{equation}
\begin{equation}
f(p_{\mu})=\int d^{4}x^{\mu}
\sqrt{-\text{det}g_{\mu\nu}(\varepsilon)}e^{ip_{\mu}x^{\mu}}f(x^{\mu})\,
.
\end{equation}
Since we are considering Fermion, the Dirac matrix $\gamma^{\mu}$
should also be parameterized to satisfy
\begin{eqnarray}
\begin{split}
\{\gamma^{\mu}(\varepsilon),\gamma^{\nu}(\varepsilon)\} &
= 2g^{\mu\nu}(\varepsilon) \, , \\
\{\gamma_{\mu}(\varepsilon),\gamma_{\nu}(\varepsilon)\} &
= 2g_{\mu\nu}(\varepsilon) \, , \\
\{\gamma^{\mu}(\varepsilon),\gamma_{\nu}(\varepsilon)\} & =
2g^{\mu}_{\nu}(\varepsilon)=2\delta^{\mu}_{\nu}\, .
\end{split}
\end{eqnarray}
With the parameterized metric, the electron self-energy at the
one-loop level can be written as
\begin{equation}
\Sigma_2(p) = ie^2\int\frac{d^4q}{(2\pi)^4}
\sqrt{-\text{det}g^{ab}}\, \gamma^\mu
\frac{-i\slashed{q}+m_e}{q^2+m_e^2}  \gamma^\nu
\frac{g_{\mu\nu}}{k^2-\mu^2} \, ,
\end{equation}
where
\begin{eqnarray}
\begin{split}
\slashed{q} & = g_{\alpha\beta} \gamma^\alpha q^\beta \, , \\
\label{p2} q^2 & = g_{\alpha\beta} q^\alpha q^\beta \, , \\
k^2 & = g_{\alpha\beta} k^\alpha k^\beta \, .
\end{split}
\end{eqnarray}
Using the Pauli-Villars regularization procedure, we have
\begin{eqnarray}
\Sigma_2(p)&=&ie^2\int\frac{d^4q}{(2\pi)^4}\sqrt{-\text{det}g^{ab}}\,
\gamma^{\mu} \frac{-i\slashed{q}+m_e}{q^2+m_e^2} \gamma^\nu \notag \\
 & &\times g_{\mu\nu}\left( \frac{1}{k^2} - \frac{1}{k^2+\Lambda^2}
 \right) \, .
\end{eqnarray}
Then
\begin{eqnarray}
S_e^{-1}(p)&=&i\slashed{p}+m_e-\Sigma_2(p) \\
&=&i\slashed{p}A(p^2)+B(p^2) \, ,
\end{eqnarray}
where the functions $A$ and $B$ can be written explicitly as
\begin{eqnarray}
A(p^2)&=&1+e^2\int\frac{d^4q}{(2\pi)^4}(e^{-i\varepsilon/2})
        \frac{1}{q^2+m_e^2} \frac{2p.q}{p^2} \notag  \\
 &&\times \left( \frac{1}{k^2} - \frac{1}{k^2+\Lambda^2} \right) \, , \\
B(p^2)&=&m_e+e^2\int\frac{d^4q}{(2\pi)^4}(e^{-i\varepsilon/2})
        \frac{4m_e}{q^2+m_e^2}\notag \\
 &&\times \left( \frac{1}{k^2} - \frac{1}{k^2+\Lambda^2} \right) \,
 .
\end{eqnarray}
which can be calculated by proper numerical integral.

Note that $\Sigma_2(p)$ can also be obtained through standard
perturbative procedure, which reads
\begin{eqnarray}\label{self-energy}
\Sigma_2(p)&=&-\frac{\alpha}{4\pi}\int_0^1dx[2i(1-x)\slashed{p}+4m_e] \notag \\
 &&\times\  \text{ln}\left( \frac{(1-x)\Lambda^2}{x(1-x)p^2+xm_e^2}\right) \, .
\end{eqnarray}
From Eq.~(\ref{self-energy}), we can also extract the structure
functions $A(p^2)$ and $B(p^2)$, where $p^2$ should be understood as
complex number but not that in Eq.~(\ref{p2}).

\begin{figure}[htb]
\centering
\includegraphics[width=8.0cm]{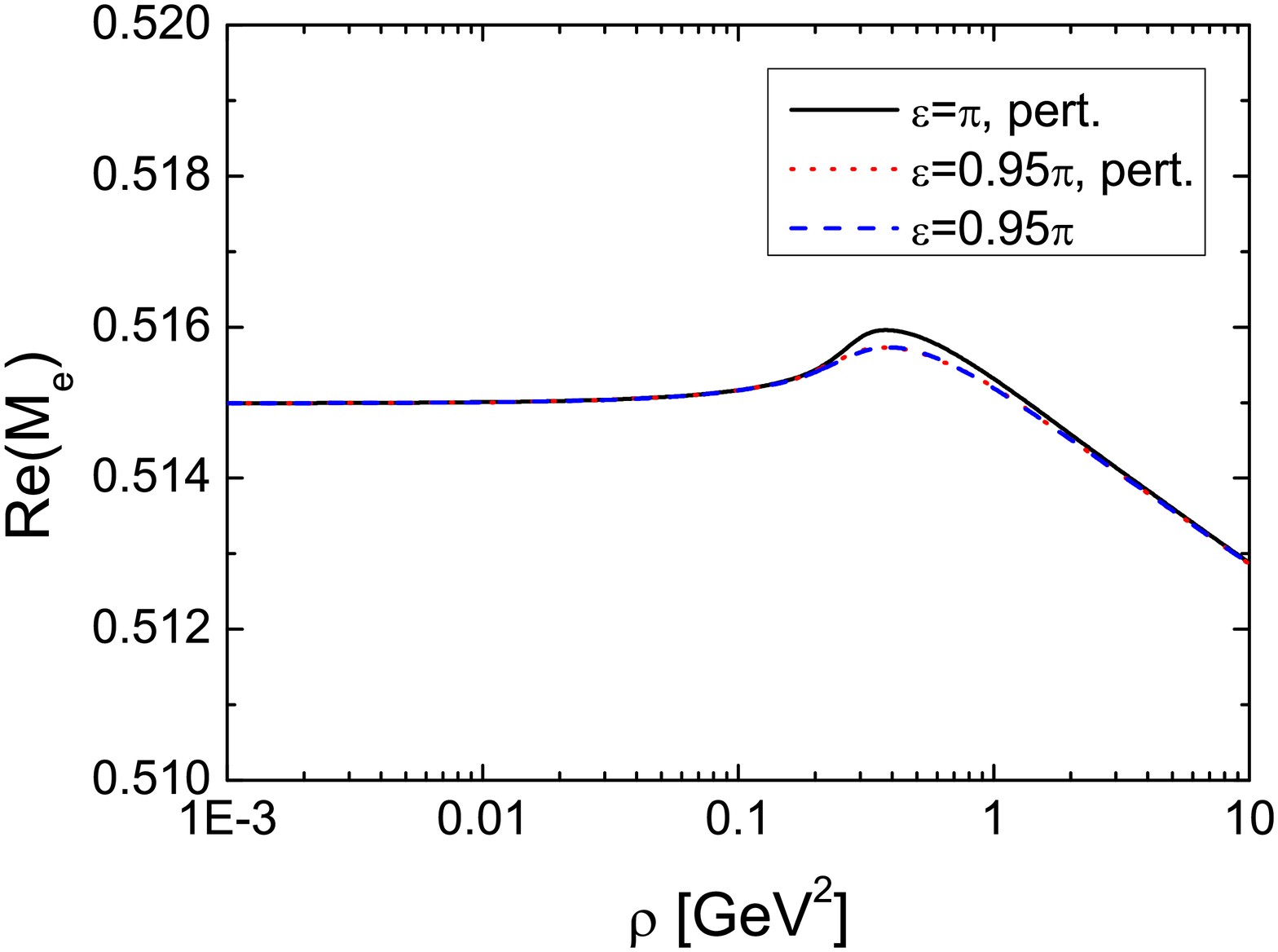}
\includegraphics[width=8.0cm]{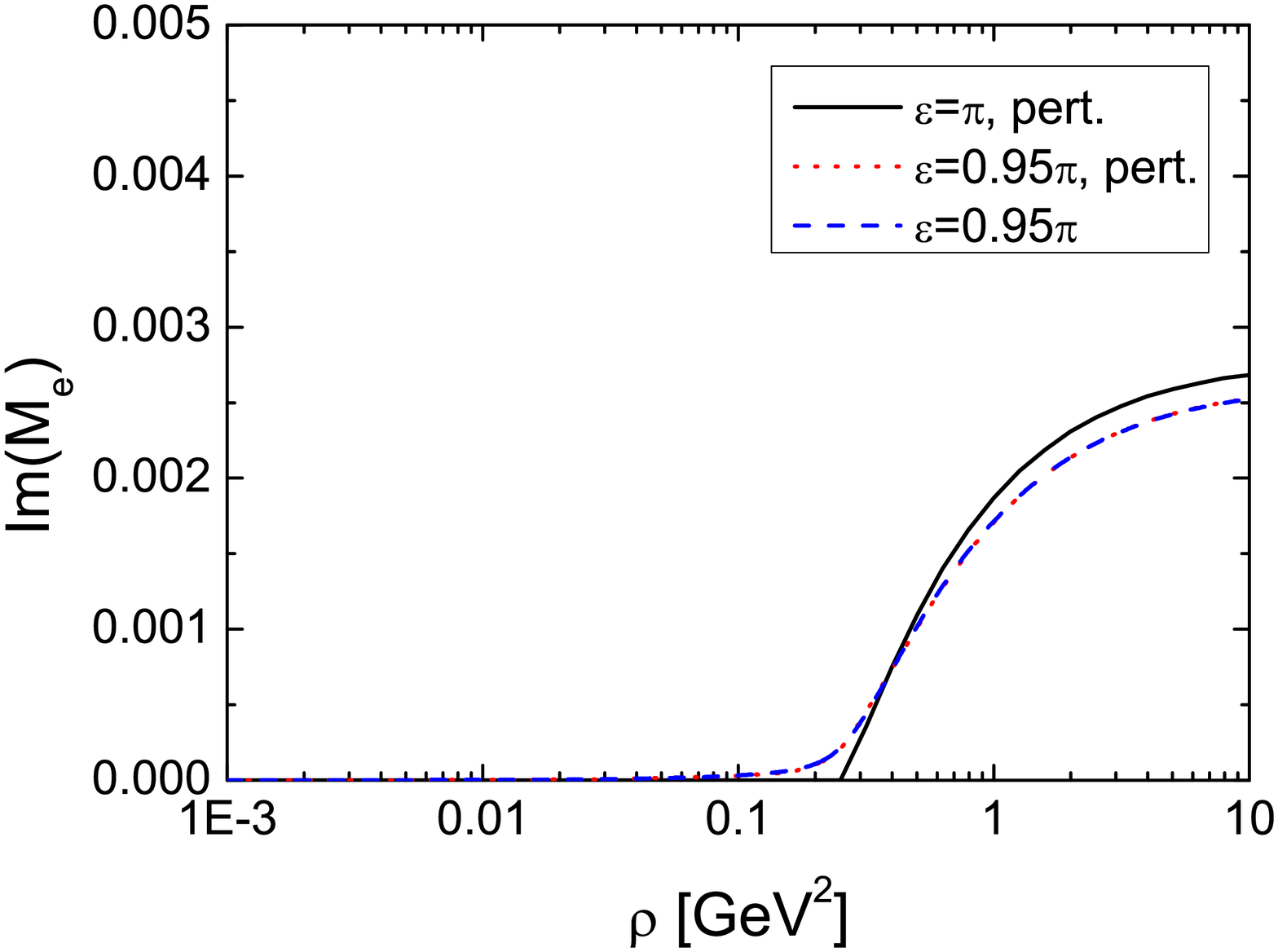}
\caption{(color online) Calculated result of the analytical
structure of the mass function in our parameterized metric and the
comparison with those given in the standard perturbative procedure,
where $p^2=\rho e^{-i\varepsilon}$. The solid line is that given by
perturbative theory in the exact Minkowskian space (with
$\varepsilon=\pi$ in the expression of $p^{2}$), the dot line is
also that calculated in perturbative theory but with
$\epsilon=0.95\pi$, the dash line is that obtained with our
parameterized metric at $\varepsilon=0.95\pi$. } \label{fig:qed}
\end{figure}

To illustrate the validity of the parameterized metric, we take the
analytical structure of the mass function $M(p^2)=B(p^2)/A(p^2)$ as
an example. The calculated results in our presently proposed
parameterized metric and that in the standard perturbative procedure
are displayed in Fig.\ref{fig:qed}.
Looking over Fig.\ref{fig:qed}, one can recognize definitely that
our presently proposed parameterized metric method is equivalent to
the traditional perturbative calculations.

It should be noted that, in this example, the equivalence between our
newly proposed method and the standard perturbative calculation is
somewhat trivial and our method is not economic. However, in
non-perturbative frameworks, traditional analytic continuation does
not work anymore (it requires us knowing the strict analytical
expression of the solutions), instead, our method will become
valuable and efficient. Anyway, such an equivalence indicates that
presently proposed parametrized metric method is absolutely valid.
Then, as we will see below, we can successfully apply this method to
Dyson-Schwinger equation. However, before we do that, there is
another problem deserving our attention, namely the vacuum problem,
because it will influence the proper truncation of Dyson-Schwinger
equation.

\section{Vacuum problem}

To demonstrate the vacuum problem clearly, we first consider a free
quark system with finite mass and chemical potential. In this case,
the analytical expression for quark propagator is exactly known. The
total influence of chemical potential $\mu$ on the quark propagator
is just a complex continuation of $p_{0}^{}$ ($p_{0}^{}\rightarrow
p_0-i\mu$). Actually, in many articles, $\mu$ is used as a single
parameter for analytic continuation and not referred to as chemical
potential. When $\mu>m$, Fermi-surface will emerge from the origin
of momentum space, and the vacuum will be changed. However, since
there is no interaction, the influence of the chemical potential
$\mu>m$ on the vacuum can not alter the quark propagator's form.
Now, we turn on some interaction which needs not to be strong (see,
for example, Refs.~\cite{Chen2008,tem1,Lei}). In this case, the
quark propagator will include self-energy part which is dependent on
the vacuum. Therefore, as the vacuum has been altered, the quark
propagator with $\mu>m$ is no longer an analytically continued Green
function with $p_0\rightarrow p_0-i\mu$. Up to now, we have not
questioned the meaning of analytic continuation. We could just say
that, beyond the mass threshold, chemical potential is no longer a
valid parameter for the task of the analytic
continuation~\cite{Chen2008}.

However, in reality, there might be even worse situation when we try
to use the parameterized metric $g_{\mu\nu}(\varepsilon)$ to realize
the analytic continuation. As we will see, with the increasing of
the parameter $\varepsilon$, the range of
$p^{2}=p_{\mu}p_{\nu}g^{\mu\nu}$ ($p_{\mu}\in R$) will become larger
and larger in the complex $p^{2}$ plane. Considering the Minkowskian
limit $\varepsilon\rightarrow\pi$, the maximal range of the complex
$p^{2}$ would be in the lower half-plane except the time-like real
axis. Once a pole enters the range of $p^{2}$, the situation becomes
very similar to what happened in the case of large chemical
potential $\mu>m$. The vacuum will be changed from then on. Because
of the interaction, Green function is no longer an analytic function
of $p^{2}$. In this case, how could we determine the physical one
na\"{i}vely?
Can we still say that we have chosen a wrong way of analytic
continuation as the same as that for chemical potential? In the
sense of analytic continuation, we are indeed wrong. However, at the
same time, we also see that the results of analytic continuation (if
exists) is no longer corresponding to the true vacuum structure
implied by the Minkowskian metric. It will result in a serious
contradiction. On one hand, choosing the strict analytic
continuation means we have to abandon the Minkowskian metric which
is our original and real purpose! On the other hand, choosing a
complete Minkowskian framework, we will lose the analyticity of
Green function! Obviously, the root of this problem is the presence
of singularities outside of the time-like real axis of $p^{2}$
plane. Actually, in this case, the Minkowskian vacuum is different
from the Euclidean one. For example, if the quark propagator has a
pair of complex poles with $\textrm{Im}(z)\neq 0$ (see, for example,
Ref.~\cite{analytic2}),
\begin{equation}
S(p)=\frac{1}{i\gamma\cdot p+z}+\frac{1}{i\gamma\cdot p+z^*},
\end{equation}
the chiral quark condensate reads
\begin{equation}
-<\bar{q}q>=2N_c Z_4 \int\frac{d^3 p}{(2\pi)^3}f_{2}
(\mid\vec{p}\mid)\, , \end{equation}
with the 3-momentum distribution function
\begin{equation}
f_2(\mid\vec{p}\mid)=\frac{1}{4\pi}\int^{\infty}_{-\infty}dp_4 tr_D
S(p) \, .
\end{equation}
For our purpose, we just need to evaluate the Euclidean
$f_{2}^{E}(0)$ and the Minkowskian $f_{2}^M(0)$
\begin{equation}
f_{2}^{E}(0)=C
\int^{\infty}_{-\infty}dp_4[\frac{z}{p_4^2+z^2}+\frac{z^*}{p_4^2+(z^*)^2}]
\, ,
\end{equation}
\begin{equation}f_{2}^{M}(0)=C
\int^{\infty}_{-\infty}idp_4[\frac{z}{p_{4}^{2}-z^{2}+i0^{+}}
+\frac{z^{*}}{p_{4}^{2}-(z^{*})^{2}+i0^{+}}] \, .
\end{equation}
After some direct calculations, we obtain the difference between the
distribution functions of the condensate in the two metrics as
\begin{equation}
f_{2}^{E}(0)-f_{2}^{M}(0) = 2 \pi C \neq 0 \, .
\end{equation}
Then the chiral condensate takes definitely different values in the
two metrics.

Since $\bar{q}q$ is a gauge invariant local operator, different
condensates (or its distribution) must come from different physical
vacuums. It is easy for us to accept a different vacuum induced by
the chemical potential $\mu > m $. However, in the framework of path
integrals in quantum field theory, the vacuums implied by the
Minkowskian metric ($\varepsilon \rightarrow \pi$) and the Euclidean
metric must be the same as we have already seen in section II. So,
if a model is well defined, the contradiction mentioned above should
not exist. It means that the singularities of the propagator can not
exist outside of the time-like real axis of $p^{2}$. With this
restriction, the parameterized metric can help us to perform the
analytic continuation successfully and will indeed preserve the
vacuum until $g^{\mu\nu}(\varepsilon)$ approaches infinitely the
Minkowskian metric.

In the following, we will try to establish such an approach in the
framework of the Dyson-Schwinger equations~\cite{DSE}.

\section{Quark's DSE models with the parameterized metric}

In this paper, the quantity of our interests is the quark propagator
in momentum space
\begin{equation}
S(p)=\int
d^4x\sqrt{-\text{det}g_{\mu\nu}}e^{ip.x}\langle\Omega|Tq(x)\bar{q}(0)|
\Omega\rangle \, .
\end{equation}
It satisfies the quark Dyson-Schwinger equation~\cite{DSE}
\begin{equation}
S^{-1}(p)=(i\gamma.p+m)+\Sigma(p)\, ,
\end{equation}
with
\begin{equation}
\Sigma(p)=\int^\Lambda_q
\sqrt{-\text{det}g^{\mu\nu}}g^2D^{\mu\nu}(k)\frac{\lambda^a}{2}\gamma_\mu
S(q)\Gamma^a_\nu(q,p)\, .
\end{equation}
And according to its Lorentz structure, it is usually decomposed as
\begin{equation}
S^{-1}(p) = i\gamma.pA(p^2)+B(p^2) \, ,
\end{equation}
or
\begin{equation}
S(p) = i\gamma.p \, \sigma_{A}^{}(p^2) + \sigma_{B}^{}(p^2) \, .
\end{equation}

Since we will make use of a model that is free of ultraviolet
divergence, here, we need not to discuss the issues related to the
regularization and renormalization of Dyson-Schwinger equation
\cite{renormalization,renormalization2}. Anyway, if one adopts a
model that has ultraviolet divergence, Pauli-Villars regularization
would be a reasonable choice just as we have done in section II.

Because the quark DSE involves effective (full) gluon propagator and
quark-gluon interaction vertex which can not be determined by quark
DSE itself, we need to choose some models for them. In Landau gauge,
we have
\begin{equation}
D^{\mu\nu}(k)=D(k)[g^{\mu\nu}(\varepsilon) - \frac{k^\mu
k^\nu}{k^2}] \, ,
\end{equation}
with $k^{\mu} \equiv g^{\mu\nu} k_{\nu}.$ And for the vertex
$\Gamma^a_\mu$, Abelian assumption gives
\begin{eqnarray}
\Gamma^{a}_{\mu} (q,p) = \frac{\lambda^{a}}{2}\Gamma_{\mu}(q,p)  \,
.
\end{eqnarray}
Finally, we need to choose some proper models for $D(k)$ and
$\Gamma_{\mu}(q,p)$. In this paper, to preserve the vacuum in the
evolution of the metric from Euclidean to Minkowskian, we have to
adopt a self-consistent gluon propagator whose singularities do not
appear outside of the time-like $k^{2} < 0$ axis. As a simple
choice, we take
\begin{equation}
g^{2}D(k)=4\pi^2D\frac{\chi^2}{(k^{2}+\Delta)^{2}} \, ,
\end{equation}
in this paper. Since gluon is confined in reality, one might have
instinctive suspicion on our choice in sense of confinement.
Nevertheless, the lack of {\it K\"{a}ll\'{e}n-Lehmann} spectral
representation is not the only way to exclude a particle from the
physical Hillbert space. For example, in QED, the longitudinal
photon indeed have legitimate {\it K\"{a}ll\'{e}n-Lehmann} spectral
representation, but it is still located outside of the physical
Hillbert space. Of course, we are far away from the solution of QCD
confinement~\cite{confinement,confinement2,confinement3}. People
need to make much more efforts for this problem. More or less, it
has gone beyond the main scope of our current discussions. Anyway,
we will come back to this issue later.

For the quark-gluon interaction vertex, the simplest choice is the
so called rainbow approximation
\begin{equation}
\Gamma^{\mu} = \gamma^{\mu} \, .
\end{equation}
In many practical calculations, the na\"{i}ve approximation has got
triumphs. However, given the Abelian assumption, the bare
quark-gluon interaction vertex violates the Ward-Takahashi identity
(WTI) which is the consequence of the conservation of quark current.
During a single-quark process, current's conservation is crucial for
the unitary. Therefore, rainbow approximation could not promise that
singularities of quark propagator can only appear at the time-like
real axis ($p^{2} < 0 $).

Beyond the bare (or rainbow) approximation, people have developed
other vertex forms that satisfy the WTI. It has been shown that the
so-called Ball-Chiu (BC) ans\"{a}tz for the vertex~\cite{BC} is a
successful form of them (see, for example, Ref.~\cite{Lei09}). It is
expressed as
\begin{equation}
\Gamma^{BC}_{\mu} = \Sigma_{A} \gamma_{\mu} +
\frac{\gamma\cdot(p+q)}{2} \Delta_{A} (p+q)_{\mu} +i \Delta_{B}
(p+q)_{\mu} \, ,
\end{equation}
with
\begin{eqnarray}
\Sigma_{A} & = & \frac{A(p^2)+A(q^2)}{2}\, ,\\
\Delta_{A} & = & \frac{A(p^2)-A(p^2)}{p^2-q^2}\, ,\\
\Delta_{B} & = & \frac{B(p^2)-B(p^2)}{p^2-q^2}\, .
\end{eqnarray}
Since BC vertex fully satisfies WTI, we expect that singularities of
quark propagator could only appear at the time-like real axis
($p^2<0$) (similar conclusion has already been drawn in
Ref.~\cite{analytic1}). Then, we will take BC vertex in our
practical calculation.

\section{Numerical Calculation and Results}

\subsection{Algorithm and its validity}

For numerical calculation, we extract at first the equations for the
functions $A$ and $B$ from the DSE of quark. After some derivation,
we can write the equations about the $A$ and $B$ in the parametrized
metric as
\begin{equation}
A(p^2)= 1 \! +\!\! \int\!\! e^{-\frac{i\varepsilon}{2}}
dq_0^2d\mathbf{q}^2 \frac{|\mathbf{q}|}{4|q_0|}\
        \frac{\Theta_A[A(p^2),A(q^2),p^2,q^2]}{q^2A^2(q^2)+B^2(q^2)}\, ,
\end{equation}
\begin{equation}
B(p^2) = \! \int\!\! e^{-\frac{i\varepsilon}{2}} dq_0^2d\mathbf{q}^2
\frac{|\mathbf{q}|}{4|q_0|}\
        \frac{\Theta_B[A(p^2),A(q^2),p^2,q^2]}{q^2A^2(q^2)+B^2(q^2)}\,
        ,
\end{equation}
with
\begin{eqnarray}
\Theta_{A} & = & A(q^2)\Sigma_{A}F^{1}_{A} + A(q^2)\Delta_{A}
F^{2}_{A}
+ B(q^2)\Delta_{B} F^{3}_{A} \, , \notag\\
\Theta_{B} & = & B(q^2)\Sigma_{A} F^{1}_{B} + B(q^2)\Delta_{A}
F^{2}_{B} + A(q^2)\Delta_{B}F^{3}_{B} \, , \notag
\end{eqnarray}
where $q^2=q_0^2e^{-i\varepsilon}+\mathbf{q}^2$, and
$F^i_{A,B}(i=1,2,3)$ are functions of $p^2$ and $q^2$ because the
angular integrals have been done. These simplified equations are the
starting point of our further numerical calculations.

\begin{figure}[htb]
\begin{center}
\includegraphics[width=7.0cm]{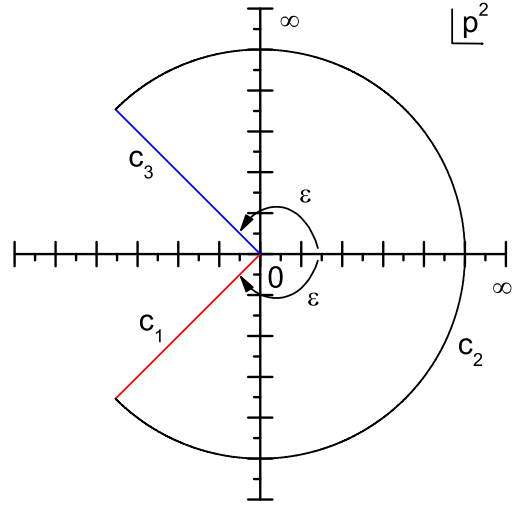}
\caption{[color online] Schematics of the area and the contour of
the complex momentum $p^2$ (the lower part) and $p^{2*}$ (the upper
part) with the parameterized metric $g^{\mu\nu}(\varepsilon)$.}
\label{fig:path}
\end{center}
\end{figure}

Note that a given metric $g^{\mu\nu}(\varepsilon)$ determines a
specific region of $p^2$ on complex plane which can be shown in
Fig.~\ref{fig:path}. Then we can parameterize $p^2$ as $\rho
e^{-i\theta}$ with $\rho\in[0,\infty)$ and
$\theta\in[0,\varepsilon]$. Obviously, when $\varepsilon$ approaches
$\pi$, the region of $p^2$ plus its conjugation will cover the whole
complex plane except the time-like axis. It should be mentioned
that, if we take the translation of $p_{0}\rightarrow p_{0}-i\mu$ to
perform the analytic continuation, the range of
$(p_{0}-i\mu)^2+\vec{p}^{2}$ is bounded by a parabolic curve. Even
if singularities only appear at time-like real axis, such parabolic
curve will inevitably encounter the singularities as $\mu$ become
moderately large. In this case, the available analytic information
is only the inner part bounded by the parabola which can not include
any singularity.
However, with the present parameterized metric, we can easily
overcome such a problem.

\newpage

With above representations, we obtain
\begin{widetext}
\begin{eqnarray}
\begin{split}
A(\rho e^{-i\theta})&=1+(e^{-i\varepsilon/2})\int_0^\infty
\rho'd\rho' \int_{0}^{\varepsilon} d\theta' \
\left(\frac{1}{4sin{\varepsilon}}\sqrt{\frac{sin(\varepsilon-\theta')}{sin\theta'}}\right)\
        \frac{\Theta_A[A(\rho e^{-i\theta}),A(\rho' e^{-i\theta'}),\rho e^{-i\theta},\rho' e^{-i\theta'}]}
        {(\rho' e^{-i\theta'})A^2(\rho' e^{-i\theta'})+B^2(\rho' e^{-i\theta'})} \, , \\
B(\rho e^{-i\theta})&=(e^{-i\varepsilon/2})\int_0^\infty \rho'd\rho'
\int_{0}^{\varepsilon} d\theta' \
\left(\frac{1}{4sin{\varepsilon}}\sqrt{\frac{sin(\varepsilon-\theta')}{sin\theta'}}\right)\
        \frac{\Theta_B[A(\rho e^{-i\theta}),A(\rho' e^{-i\theta'}),\rho e^{-i\theta},\rho' e^{-i\theta'}]}
        {(\rho' e^{-i\theta'})A^2(\rho' e^{-i\theta'})+B^2(\rho'
        e^{-i\theta'})}\, .
\end{split}
\end{eqnarray}
\end{widetext}

In principle, the above equations can be solved directly. In
practice, to simplify numerical calculations, we consider
ultraviolet boundary conditions for the functions $A$ and $B$ and
the Cauchy integral formulas~\cite{analytic2} which read
\begin{eqnarray}
&&A(C_{3})=A^*(C_{1})\, , \quad B(C_{3})=B^*(C_{1})\, , \\
&&A(C_{2})=1\, , \qquad \quad \; \; \, B(C_{2})=0 \, , \\
&&F(z)=\frac{1}{2\pi i}\oint_{C_{1} + C_{2} + C_{3}}
\frac{F(t)}{t-z}dt \, ,
\end{eqnarray}
where the paths $C_{1}$, $C_{2}$ and $C_{3}$ are those shown in
Fig.~\ref{fig:path}, and $F$ stands for the functions $A$ and $B$.
Then, in the integral equations, the only unknown function is
$F(C_1)$, or $F(\rho e^{-i\varepsilon})$ which can be fully solved.
Therefore, the original two-dimensional problem is reduced to a
one-dimensional one. Such a reduction is very significant because it
makes the numerical calculation with large $\varepsilon$
($\rightarrow \pi$) possible.

\begin{figure}[htb]
\centering
\includegraphics[width=8.0cm]{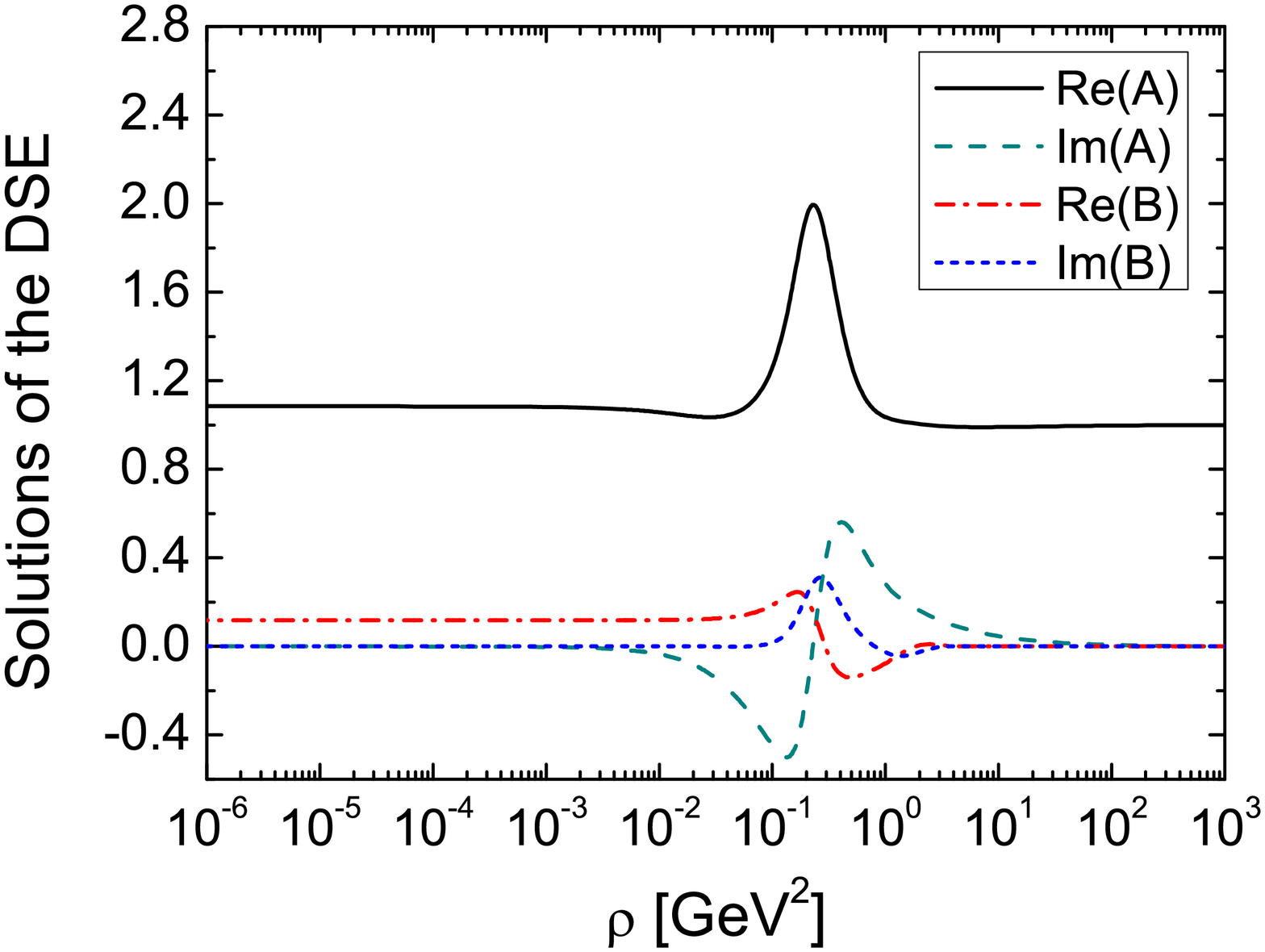}
\caption{[color online] Calculated variation behaviors of $A(\rho
e^{-i\varepsilon})$ and $B(\rho e^{-i\varepsilon})$ with respect to
$\rho$ at $\varepsilon=3\pi/4$.}\label{fig:3pi-4}
\end{figure}

\begin{figure}[htb]
\centering
\includegraphics[width=8.0cm]{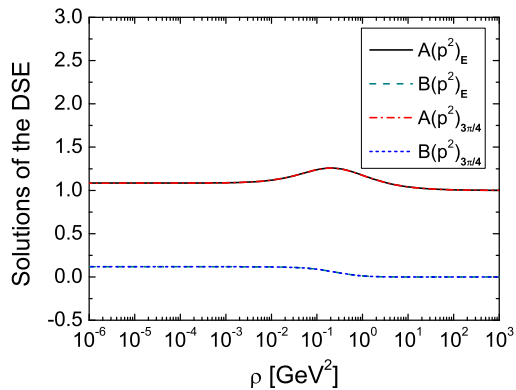}
\caption{[color online] The comparison between the data generated by
our algorithm and those obtained directly in the Euclidean
space.}\label{fig:comp}
\end{figure}

To check the validity of our method, we adopt firstly a relatively
small $\varepsilon=3\pi/4$ to solve the equation (with
$D=0.25\text{GeV}^2$, $\chi=1.0\text{GeV}^2$ and
$\Delta=0.1\text{GeV}^2$). The solutions of the functions $A$ and
$B$ are illustrated in Fig.~\ref{fig:3pi-4}. Making use of these
solutions, we can generate the data at $\theta =0$ with the Cauchy
integral formulas.
We compare then the generated data with the solutions obtained by
solving directly the equations in Euclidean space (solving the
equations after taking $\varepsilon = 0$). The comparison is
displayed in Fig.~\ref{fig:comp}. The figure manifests apparently
that the two groups of curves are exactly coincident. It indicates
that the validity of our method is in very high precision and the
vacuum really remains against the evolution of $\varepsilon$ from 0
to $0.75\pi$.

\begin{figure}[htb]
\centering
\includegraphics[width=7.50cm]{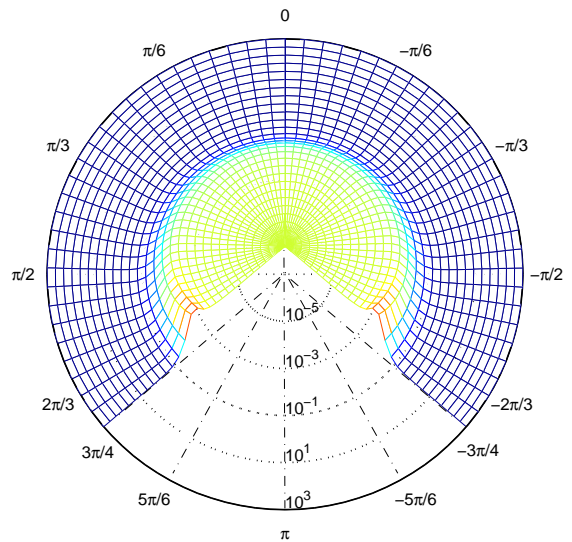}
\includegraphics[width=8.0cm]{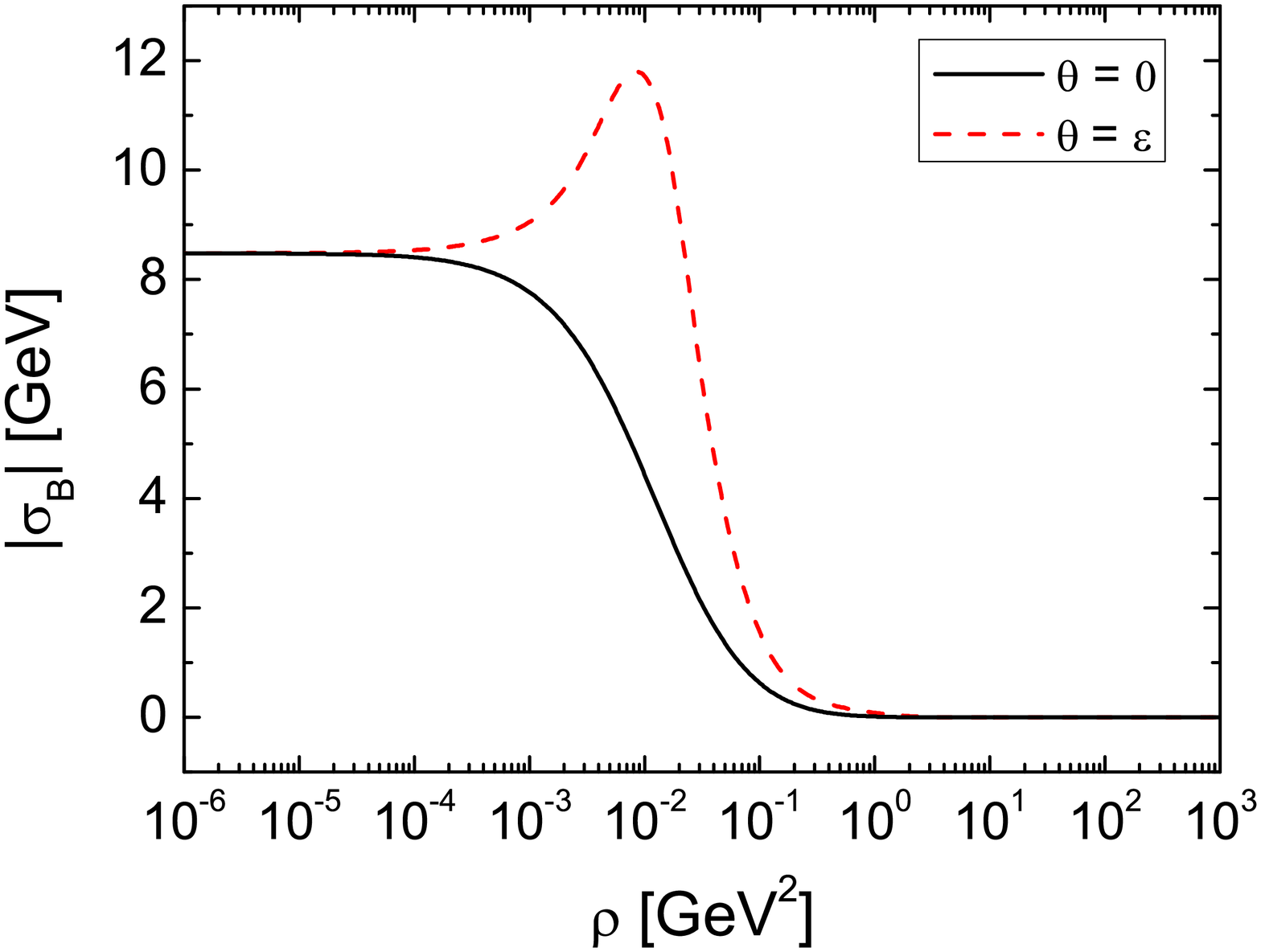}
\caption{[color online] Calculated variation behavior of the scalar
part of the quark propagator $\vert {\sigma_{B}^{}} \vert $ in the
whole momentum area in the case of $\varepsilon=3\pi/4$ (upper
panel) and that at two special values of $\theta$ (lower panel).}
\label{fig:all3pi-4}
\end{figure}

To clearly illustrate the self-consistency, we have to show that the
quark propagator has no pole in the corresponding area. We display
then the obtained quark propagator (actually the absolute value of
the $\sigma_{B}$, denoted as $\vert \sigma_{B} \vert $, is
sufficient for this checking) in the whole range of $p^{2}$ in
Fig.~\ref{fig:all3pi-4}. It is apparent that there is indeed no pole
located in the range of $p^2$.

In principle, for a given $\varepsilon$, as long as the Green
functions have no singularity in the involved momentum region, our
results would be reliable. As we have mentioned above, the
singularities of the quark propagator could only appear at the
time-like axis, then the analytic continuation on the whole $p^2$
plane except the time-like axis could be achieved by the limit
$\varepsilon \rightarrow \pi$.

\subsection{{\it Almost} Minkowskian propagator}

In this subsection, we discuss the quark propagator in the case that
is very close to the Minkowskian by performing the calculations with
increasing the $\varepsilon$ to a value highly close to $\pi$. It
should be noted that, the true limit ($\varepsilon=\pi$) is very
hard to achieve (and not necessary) in practice. At our current
level, the largest $\varepsilon$ that can yield reliable results is
$0.95\pi$. In addition, we consider the dependence of the quark
propagator on the coupling strength $D$ and the screening width
$\Delta$.

\begin{figure}[htb]
\centering
\includegraphics[width=8.0cm]{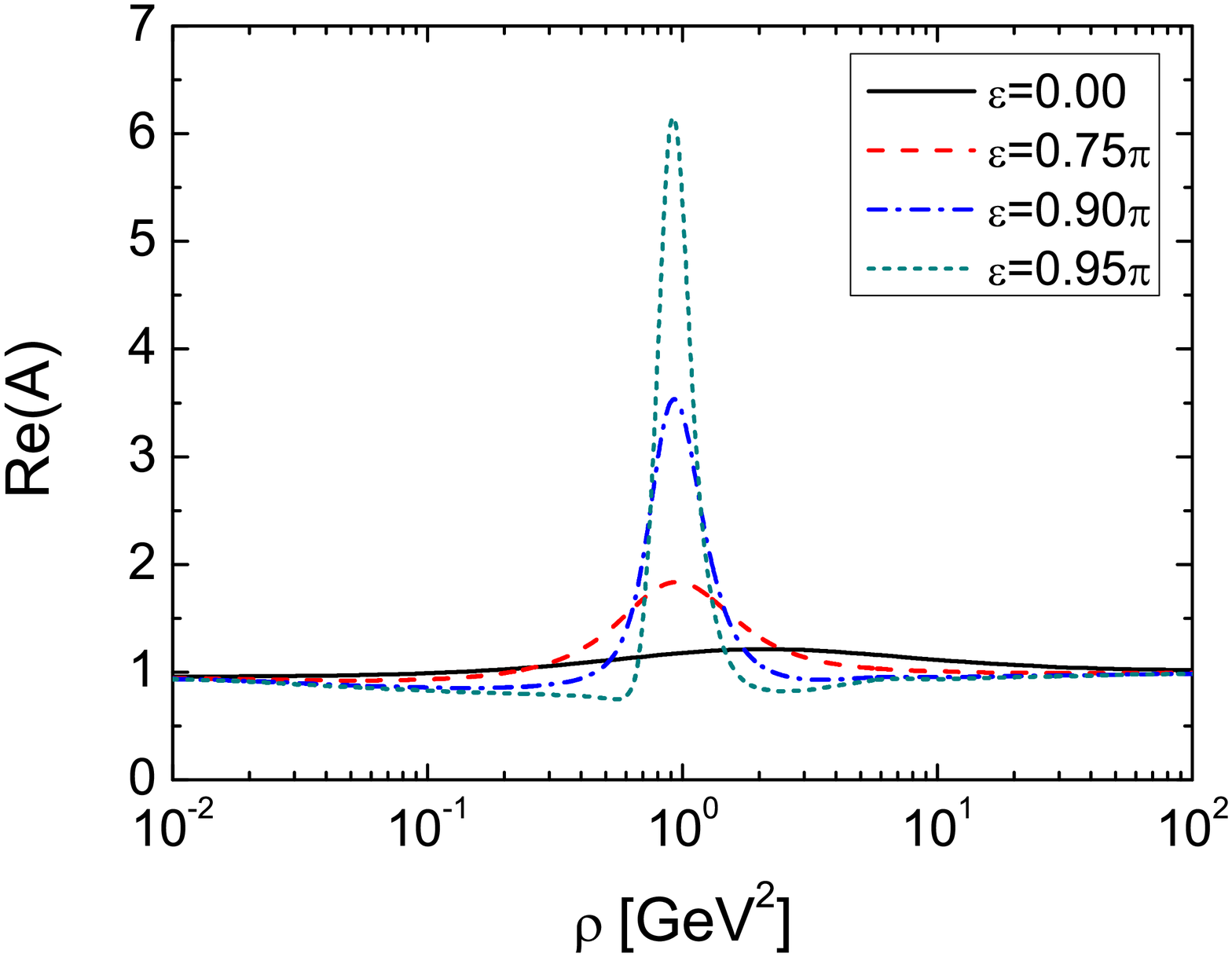}
\includegraphics[width=8.0cm]{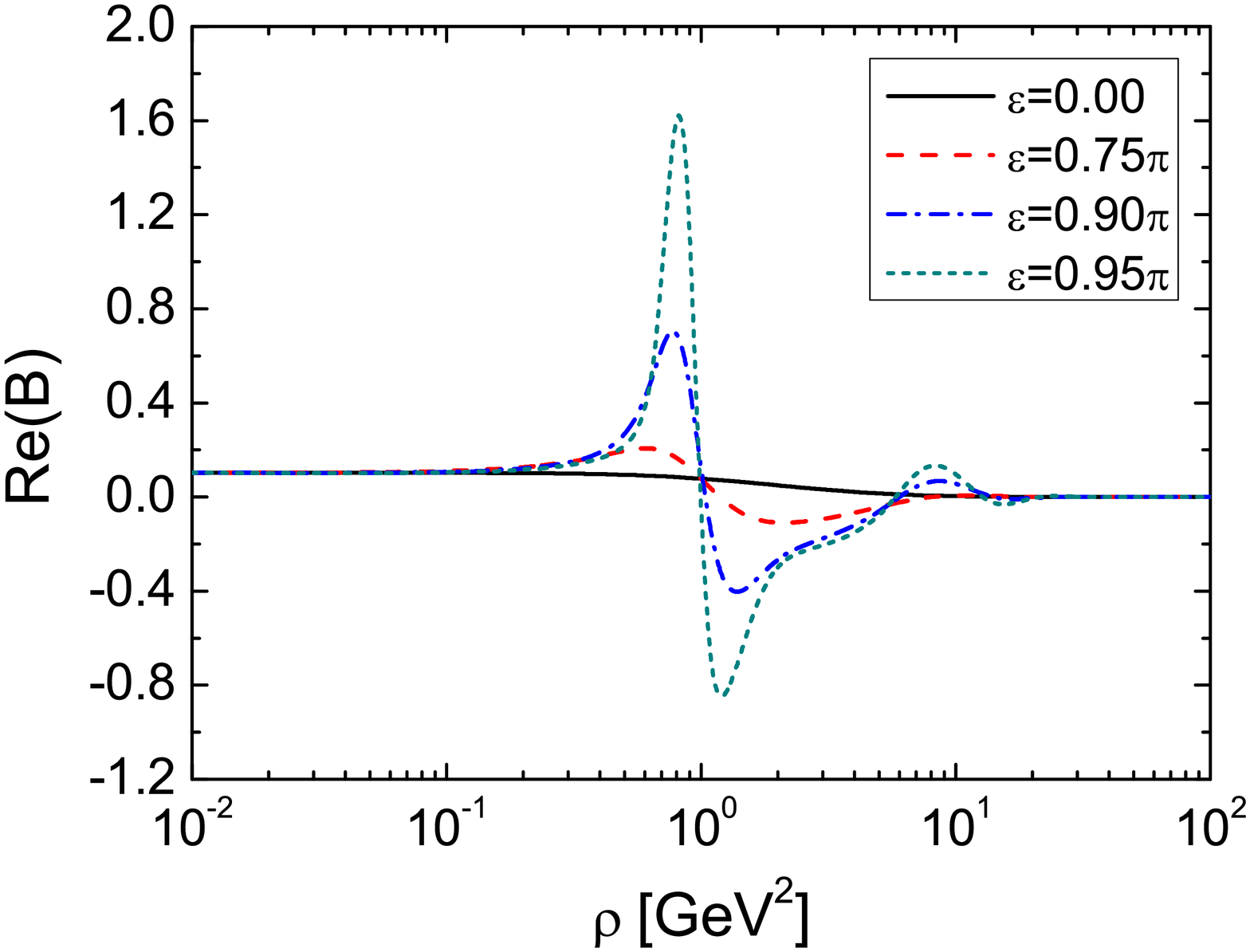}
\includegraphics[width=8.0cm]{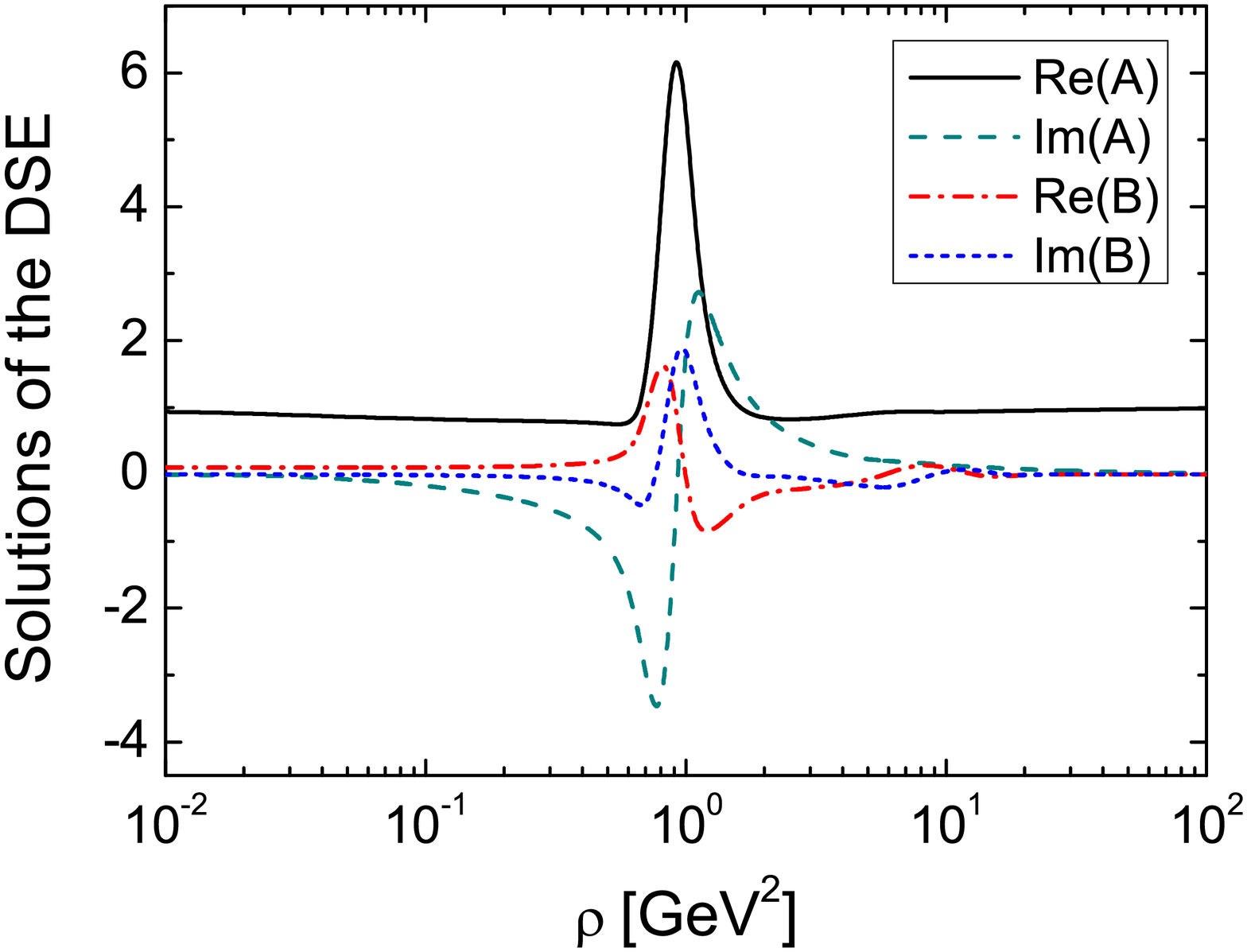}
\caption{[color online] Calculated results of the real part of the
functions $A$ and $B$ in the quark propagator ($\text{Re}(A)$ in
upper panel, $\text{Re}(B)$ in middle panel) with several values of
parameter $\varepsilon$ and those of the real and the imaginary
parts of the functions $A$ and $B$ at $\varepsilon=0.95\pi$ (lower
panel). } \label{fig:allinone}
\end{figure}

After having solved the quark DSE with our algorithm, we first
analyze the behaviors of the functions $A(p^2)$ and $B(p^2)$ in the
quark propagator against the parameter $\varepsilon$ and with fixed
parameter set $D=1.06\text{GeV}^2$, $\chi=1.0\text{GeV}^2$ and
$\Delta=0.5\text{GeV}^2$. The obtained results of the real part of
the functions $A$ and $B$ at several values of $\varepsilon$ are
illustrated in the upper, middle panel of Fig.~\ref{fig:allinone},
respectively, and both the real and imaginary parts of the $A$ and
$B$ at the maximal value of $\varepsilon$ we reached in the lower
panel of Fig.~\ref{fig:allinone}. We can notice apparently from the
figure that both the functions $A$ and $B$ have a peak at some
complex momentum. Moveover, the peaks grow sharper and sharper with
the increasing of $\varepsilon$. Although we have to stop at
$\varepsilon=0.95\pi$ because of numerical difficulty (the peaks
make the numerical contour integral inexact as the $\varepsilon$
approaches much more to $\pi$), we can predict that the peaks will
finally evolve into singularities. Namely, the analytic functions
$A(p^2)$ and $B(p^2)$ both have poles at $p^{2}_{g}$ (on the
time-like momentum axis).

To show the limit behavior of the imaginary part of $A$ and $B$ as
$\varepsilon\rightarrow \pi$, we need the results with $\varepsilon$
much more close to $\pi$. Up to now, we can not say any definite
things about the branch-cut structure of the quark propagator yet.
However, there are some indications that the power of the divergence
of $A$ and $B$ is less than one and there would exist branch-cut on
the time-like axis. For instance, we find that the absolute values
of $A $ and $ B $ involve oscillation when $p^2<p^{2}_{g}$ which
seems to come from logarithmic divergence. Another evidence is the
symmetric behavior shown in Fig.~\ref{fig:all3pi-4}. Since
Fig.~\ref{fig:all3pi-4} is drawn in $\ln{p^{2}}$, the symmetric
feature around the pole also supports logarithmic divergence.
However, this issue needs further investigation.

\begin{figure}[htb]
\centering
\includegraphics[width=7.50cm]{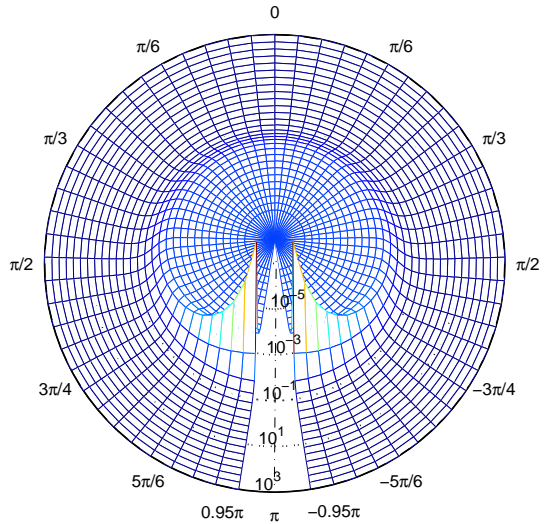}
\caption{[color online] Calculated result of the variation behavior
of the function $\vert {\sigma_{B}^{}} \vert$ in the whole momentum
area in the case of $\varepsilon=0.95\pi$.}\label{fig:m950}
\end{figure}

Then, from the behaviors of the functions $A$ and $B$, we can obtain
the quark propagator. The obtained results of the $\vert
{\sigma_{B}^{}} \vert$ in the whole momentum region with the
reachable maximal $\varepsilon$ ($=0.95\pi$) is shown in
Fig.~\ref{fig:m950}. One could easily find from the figure that the
Minkowskian quark propagator has one pole at $p^{2}_{q}$, one
zero-point at $p^{2}_{g}$, respectively. Noting that both the $A$
and $B$ are approximately constant at $p^{2}>p^{2}_{g}$, which means
that the dynamical mass function $M(p^{2})=B(p^{2})/A(p^{2})$ also
roughly remains constant, one can then recognize that, as the
time-like momentum $p^{2} \approx -M^{2}(0)>p^{2}_{g}$, the
structure $1/[p^{2} + M^{2}(p^{2})]$ will generate a simple pole.
Obviously, the location of the pole is roughly determined by the
dynamical mass at zero momentum, namely, $p^{2}_{q} \approx
-M^{2}(0)$. For the zero-point, it comes evidently from the
divergence of the $A$ and $B$ at $p^{2}_{g}$.

It should be emphasized that the simple pole of the Minkowskian
quark propagator will likely lead to an unconfined quark. Although
the result is explicitly against the property of confinement, we
tend to think that it is caused by the specific model for the gluon
propagator. Actually, the gluon model adopted in this paper is not a
rigorously confined one, because the linear confinement potential
requires $\Delta \rightarrow 0$. Then, we wonder what will happen
when we decrease the parameter $\Delta$ and maintain the mass gap
$M(0)$. The obtained results at several values of the $\Delta$ are
presented in Table~\ref{tab:parad}. It is evident that $p^{2}_{g}$
approaches $p^{2}_{q}$ with the decreasing of $\Delta$. It is a
positive signal for our assumption that $\Delta \rightarrow 0$ will
lead to $p^{2}_{g} \rightarrow p^{2}_{q}$. However, we could not
prove this statement yet, since numerical difficulty prevents us
from decreasing the $\Delta$ to infinitesimal. In this article, we
leave this question open. However, given such an assumption, we
could expect that, as $p^{2}_{g} \rightarrow p^{2}_{q}$, the simple
pole and the zero point of the quark propagator will combine into a
lower-order pole (fractional power or logarithmic divergence) which
finally confines quark.
\begin{table}[htb]
\caption{Calculated results of the parameter dependence of the
Minkowskian quark propagator} \label{tab:parad}
\begin{center}
\begin{tabular}{|c|c|c|c|c|}
\hline
$D$~(GeV$^2$) & $\Delta$~(GeV$^2$) & $p^{2}_{g}$~(GeV$^2$) & $M(0)$~(GeV) & $p^{2}_{q}$~(GeV$^2$) \\
\hline
1.06 & 0.50 & $-0.88$ & 0.11 & $-0.012$ \\
0.66 & 0.30 & $-0.58$ & 0.11 & $-0.012$ \\
0.25 & 0.10 & $-0.22$ & 0.11 & $-0.012$ \\
\hline
\end{tabular}
\end{center}
\end{table}

\section{Summary and Remarks}

When we have interests with the quantities, such as conductance and
viscosity, which involve transport processes, the Minkowskian metric
would be inevitable. The traditional method
 to realize the Minkowskian metric is the so-called analytic continuation of Green function.
In this paper, we argue that the meaningful analytic continuation of
Green functions requires an identical vacuum. It should be true for
any rigorous quantum field theory, but not always promised by
models. Besides this problem, the widely used method for analytic
continuation in various models is based on the traditional {\it
K\"{a}ll\'{e}n-Lehmann} spectral representation. However, in some
circumstances, especially for quark fields, the situation would
become more subtle and complicated. In vacuum, since quark is
confined, quark propagator may not have standard {\it
K\"{a}ll\'{e}n-Lehmann} spectral representation. More or less, such
a difficulty may even exist in hot/dense quark matter where the
quark degree of freedom becomes important. Therefore, it will
invalidate the method of analytic continuation based on the {\it
K\"{a}ll\'{e}n-Lehmann} spectral representation. It has also been
emphasized that, if we implement the translation $p_{0}\rightarrow
p_{0}-i\mu$ to perform the analytic continuation, the range of
$(p_{0}-i\mu)^2+\vec{p}^{2}$ is bounded by a parabolic curve. Even
if singularities only appear at time-like real axis, such parabolic
curve will inevitably encounter the singularities as $\mu$ become
moderately large. In this case, the available analytic information
is only that in the inner part of the parabola whose top approaches
the first singularity.

In this paper, we have developed a new scheme which is free of all
the problems mentioned above. A well-behavior effective gluon
propagator and the Ball-Chiu vertex ans\"{a}tz can promise that the
singularities can only appear at time-like axis, thus promise an
uniform vacuum under Euclidean and Minkowskian metrics. Furthermore,
by establishing a parameterized metric $g^{\mu \mu}(\varepsilon)$
which connects the Euclidean metric with the Minkowskian one, we can
directly solve the Green function in the full $p^2$ plane except the
time-like axis without implementing any priori decomposition rules
such as the {\it K\"{a}ll\'{e}n-Lehmann} spectral representation.
After confirming the correctness of our parameterized metric scheme
in the simple $\phi^{4}$ model and in the calculation of electron
self-energy at one-loop level and checking the validity of the
algorithms for solving the quark DSE in the case of a relative small
parameter $\varepsilon$ ($=0.75\pi$), we continuously increase the
$\varepsilon$ and obtain the almost Minkowskian quark propagator
with $\varepsilon=0.95\pi$. We find that the Minkowskian quark
propagator has one simple pole and one zero-point at $p^2_q$ and
$p^2_g$ respectively. By observation on the dependence of the quark
propagator on the screening width $\Delta$ in the effective gluon
propagator, we could expect that, as $\Delta \rightarrow 0$, the
simple pole and the zero point of quark propagator will combine into
a lower-order pole (fractional power or logarithmic divergence)
which finally confines quark. However, the exact structure (for
instance, branch-cut) has not yet been identified due to the limited
$\varepsilon$. In hot/dense strong interaction matter where the
quark degree of freedom becomes more significant and
practical~\cite{sQGP,Wang2007,Chen2008,tem1,tem2,tem3}, our method
need further modifications which is under way. In this case, an
interesting question is whether the so-called deconfined quark has
the standard {\it K\"{a}ll\'{e}n-Lehmann} spectral representation.
Anyway, an opposite answer can only come from the non-perturbative
nature.

In addition, if one makes use of a gluon propagator model, whose
ultraviolet behavior being consistent with the perturbative result,
the loop integral in quark DSE will be divergent. In this case,
proper regularization would be inevitable. To be honest, we have no
idea about how to take dimensional regularization in the numerical
calculation at present stage. Since $p^2$ is a complex number in the
calculation, the regularization set by $p^2<\Lambda^2$ is also
invalid. One might take $|p^2|<\Lambda^2$ alternatively. However, it
is a non-analytic cut-off and inconsistent with our method. In our
view, Pauli-Villars regularization procedure could be a reasonable
choice, as we have taken in section II, which can overcome the
difficulties mentioned above. The related investigation is in
progress.

\section*{Acknowledgements}

This work was supported by the National Natural Science Foundation
of China under contract Nos. 10425521, 10675007 and 10935001, the
Major State Basic Research Development Program under contract Nos.
G2007CB815000.
Helpful discussions with Dr. Lei Chang are acknowledged with great
thanks.



\begin{thebibliography}{90}

\bibitem{Peskin-textbook}
  E. Peskin and V. Schoroeder,
    {\it An Introduction to Quantum Field Theory} (Westview Press, USA,
    1995).
%
%

\bibitem{sQGP}
  M. Gyulassy, and L. McLerran,
    Nucl. Phys. A {\bf 750}, \textrm{30} (2005);
  E. Shuryak,
    Nucl. Phys. A {\bf 750}, \textrm{64} (2005).
  E. Shuryak,
    Prog. Part. Nucl. Phys. {\bf 62}, \textrm{48} (2009).

\bibitem{Wang2007}
  A. Schafer, Xin-Nian Wang, and Ben-Wei Zhang,
    Nucl. Phys. A {\bf 793} \textrm{128} (2007);
  E. Shuryak,
    Prog. Part. Nucl. Phys. {\bf 62}, \textrm{48} (2009).

\bibitem{lattice}
 M. S. Bhagwat, P. C. Tandy,
   AIP Conf. Proc. {\bf 842}, \textrm{225} (2006).

\bibitem{DSE}
  C. D. Roberts, A. G. Williams,
   Prog. Part. Nucl. Phys.  {\bf 33}, \textrm{477} (1994).
 R. Alkofer, and L. von Smekal,
   Phys. Rep. {\bf 353}, 281 (2001);
 C. D. Roberts, M. S. Bhagwat, A. Hoell, S. V. Wright,
   Eur. Phys. J.- ST {\bf 140}, 53 (2007);
 C. D. Roberts,
   Prog. Part. Nucl. Phys. {\bf 61}, 50 (2008).

\bibitem{analytic1}
 R. Alkofer, W. Detmold, C. S. Fischer, P. Maris,
   Phys. Rev. D {\bf 70}, \textrm{014014} (2004).


\bibitem{Chen2008}
   Huan Chen, Wei Yuan, Lei Chang, Yu-Xin Liu, Thomas Klahn and C. D.
   Roberts,
    Phys. Rev. D {\bf 78}, \textrm{116015} (2008).

\bibitem{tem1}
 C. D. Roberts, S. M. Schmidt,
    Prog. Part. Nucl. Phys.  {\bf 45}, \textrm{S1} (2000).
 P. Maris, and C. D. Roberts,
   Int. J. Mod. Phys. {\bf 12}, 297 (2003);

\bibitem{Lei}
  H.S. Zong, L. Chang, F.Y. Hou, W.M. Sun, and  Y.X. Liu,
    Phys. Rev. C {\bf 71}, 015205 (2005).
  F.Y. Hou, L. Chang, W.M. Sun, H.S. Zong, and  Y.X. Liu,
    Phys. Rev. C {\bf 72}, 034901 (2005).

\bibitem{analytic2}
  N. I. Ioakimidis, K. E. Papadakis, E. A. Perdios,
    BIT  {\bf 31}, \textrm{276} (1991).


\bibitem{renormalization}
  C. S. Fischer, R. Alkofer,
    Phys. Rev. D {\bf 67}, \textrm{094020} (2003).

\bibitem{renormalization2}
  D. C. Curtis, M. R. Pennington,
    Phys. Rev. D {\bf 42}, \textrm{4165} (1990).


\bibitem{confinement}
  T. Kugo,
    arXiv: \textit{hep-th/95511033};

\bibitem{confinement2}
  D. Zwanziger,
    Nucl. Phys. B {\bf 364}, \textrm{127} (1991).


\bibitem{confinement3}
  R. Alkofer, C. S. Fischer, L. V. Smekal,
    Prog. Part. Nucl. Phys. {\bf 50}, \textrm{317} (2003).


\bibitem{BC}
  J. S. Ball, T.-W. Chiu,
    Phys. Rev. D {\bf 22}, \textrm{2550} (1980).

\bibitem{Lei09} L. Chang, and C. D. Roberts,
    Phys. Rev. Lett. {\bf 103}, \textrm{081601} (2009).

\bibitem{tem2}
  A. Bender, D. Blaschke, Y. Kalinovsky, C. D. Roberts,
    Phys. Rev. Lett. {\bf 77}, \textrm{3724} (1996).

\bibitem{tem3}
D. Blaschke, C. D. Roberts,
    Nucl. Phys. A {\bf 642}, \textrm{197} (1998).


\end{thebibliography}
\end{document}